\documentclass[12pt,a4paper]{article}
\usepackage{authblk}
\usepackage{natbib}
\usepackage{amssymb}                    % Package for mathematics
\usepackage{amsmath}
\usepackage{chngcntr}  % Change the way of cointing figure in appendix
\usepackage{booktabs}  % Horizontal line in tables
\usepackage{setspace}
\usepackage{float}
\usepackage{a4wide}
\usepackage{lineno} % Add line number
\usepackage[symbol]{footmisc}
\usepackage{lipsum}
\usepackage{rotating}
\usepackage{hyperref}

\usepackage{graphicx}
\usepackage{color}
\setkeys{Gin}{draft=false}

\title{Coupled Boundary Element and Finite Volume Methods for Modeling Fluid-Induced Seismicity in Fault Networks within Low-Permeability Rocks}

\author[1,2]{Pierre ROMANET}
\author[1]{Marco Maria SCUDERI}
\author[2]{Jean-Paul AMPUERO}
\author[3]{Stéphanie CHAILLAT}
\author[2]{Frédéric CAPPA}

\affil[1]{Department of Earth Sciences, La Sapienza University of Rome, Rome, Italy}
\affil[2]{Université Côte d’Azur, CNRS, Observatoire de la Côte d’Azur, IRD, 
Geoazur, Sophia-Antipolis, France, email: romanet@geoazur.unice.fr}
\affil[3]{Laboratoire POEMS, CNRS-INRIA-ENSTA Paris, Institut Polytechnique de Paris}

\begin{document}
\maketitle
\doublespacing

%----------------------------------------------------------------------------------------
%	TITLE
%---------------------------------------------------------------------------------------- y 

%----------------------------------------------------------------------------------------
%	AUTHORS AND AFFILIATIONS
%----------------------------------------------------------------------------------------

% Use \author{\altaffilmark{}} and \altaffiltext{}

% \altaffilmark will produce footnote; matching \altaffiltext will appear at bottom of page.

%----------------------------------------------------------------------------------------
%	ABSTRACT
%----------------------------------------------------------------------------------------

% Do NOT include any \begin...\end commands within the body of the abstract.b

\begin{abstract}
To better understand the mechanics of injection-induced seismicity, we developed a two-dimensional numerical code to simulate both seismic and aseismic slip on non-planar faults and fault networks driven by fluid diffusion along permeable faults. Our approach integrates a boundary element method to model fault slip governed by rate-and-state friction with a finite volume method for simulating fluid diffusion along fault networks. We demonstrate the method’s capabilities with two illustrative examples: (1) fluid injection inducing slow slip on a primary rough, rate-strengthening fault, which subsequently triggers microseismicity on secondary, smaller faults, and (2) fluid injection on a single fault in a network of intersecting faults, leading to fluid diffusion and reactivation of slip throughout the network. In both cases, the simulated slow slip migrates more rapidly than the fluid pressure diffusion front. The observed migration patterns of microseismicity in the first example and slow slip in the second example resemble diffusion processes but involve diffusivity values that differ significantly from the fault hydraulic diffusivity. These results support the conclusion that the microseismicity front is not a direct proxy for the fluid diffusion front and cannot be used to directly infer hydraulic diffusivity, consistently with some decametric scale in-situ experiments of fault activation under controlled conditions. This work highlights the importance of distinguishing between mechanical and hydrological processes in the analysis of induced seismicity, providing a powerful tool for improving our understanding of fault behavior in response to fluid injection, in particular when a network of faults is involved.
\end{abstract}

%####################################################
%   INTRODUCTION
%##############################################
\newpage
\section{Introduction} 
    Understanding human-induced seismicity associated with anthropogenic activities has become an increasingly critical challenge for society. This urgency arises from the growing demand for energy and the need to mitigate carbon dioxide (CO2) emissions, which push for the development of green energy solutions. Many technologies developed to address these needs involve the injection of pressurized fluids into the subsurface. This is the case of Enhanced Geothermal Systems (EGS) which extract geothermal energy from deep rocks with initially low permeability, of disposal of wastewater from oil and gas operations into deep formations, and of carbon storage into sealed underground reservoirs. While these technologies provide substantial benefits in energy production and environmental management, they also introduce risks, particularly the potential of inducing earthquakes \citep{ellsworth2013,grigoli2017,keranen2018}. 
    
    In recent years, the rate of induced seismicity has increased significantly in regions with active fluid injection, leading to several relatively large earthquakes. Examples include the 2017 Mw 5.5 earthquake near an enhanced geothermal site in Pohang, South Korea \citep{grigoli2018,kim2018,lee2019}. This earthquake caused substantial damage and has been proposed to be triggered by fluid diffusion from injection activities \citep{kim2018,woo2019,ellsworth2019}. Another notable example is the 2011 Mw 5.7 and 5.0 earthquakes near Prague, in Oklahoma, USA \citep{keranen2013,sumy2014}, in which a first $M_w\;5.0$ earthquake was likely triggered by wastewater injection, and a subsequent $M_w\;5.7$ event was likely triggered by Coulomb stress changes. These events demonstrate that fluid injection can destabilize faults and trigger seismic events.
    
    Fluid injection, however, does not systematically trigger a seismic rupture \citep{cornet2012,ellsworth2013,wei2015,lengline2017}. During reservoir stimulations and field experiments, observations have shown that aseismic slip can propagate from the injection point, potentially triggering seismicity further away \citep{guglielmi2015,wei2015,cappa2019,eyre2019}. For instance, in hydraulic fracturing operations within low permeability formations containing faults, aseismic slip may occur on stable sections of a fault, which gradually load more distant, unstable regions. Eventually, this loading can cause dynamic rupture when the aseismic slip front reaches a fault segment with frictional rate-weakening behavior. This concept suggests that aseismic slip can propagate from the reservoir depth to the seismogenic region of a fault, potentially triggering earthquakes \citep{eyre2019,atkinson2020,eyre2022}. Other cases, such as those at the Soultz-sous-Forêts (France), Brawley (USA) and Cooper Basin (Australia) geothermal sites, further illustrate how fluid injection can induce aseismic slip that advances ahead of the fluid pressure diffusion front, causing shear stress perturbations that trigger seismicity \citep{danre2024,wang2022}. These cases highlight the role of fluid diffusion along critically stressed fault systems in initiating aseismic slip first, and then triggering seismicity. 
    
    In-situ experiments involving decametric fault reactivation have been essential for understanding the relationship between injected fluids, dilatancy and fault slip under controlled conditions \citep{guglielmi2015,duboeuf2017,cappa2019,deBarros2024,cappa2022a}. For instance, recent observations at the Mont-Terri underground laboratory have shown that fluid injection in a highly velocity-strengthening fault \citep{orellana2018, bigaroni2023} with very low initial permeability ($k=10^{-17} \mathrm{m}^2$) can induce significant dilatant aseismic slip, which is a key component of the deformation process in an interconnected fault network \citep{guglielmi2020,cappa2022b}. This slip is often associated with an increase in low-frequency tremor signals, linked to fluid-driven slip propagation \citep{deBarros2023}. In such fault zones, structural heterogeneity, such as variations in fracture density, mineralogy, and hydromechanical properties, plays a crucial role in controlling fluid diffusion, stress accumulation and aseismic slip. Areas with rate-strengthening friction may experience aseismic slip, while stress perturbations from aseismic slip may generate seismicity in regions with rate-weakening properties \citep{cappa2019, deBarros2023}. 
    
    Furthermore, centimeter scale laboratory experiments and decametric scale field tests have revealed the complexity of fault slip behavior with increasing fluid pressure \citep{scuderi2016b,passelegue2018,cappa2019,deBarros2023}. Fluid pressurization can induce a variety of slip modes (seismic versus aseismic) in association with fault dilation, depending on fault hydromechanical properties, stress conditions, and fluid flow dynamics. These findings emphasize the importance of understanding the relationship between fault rheology, fluid-induced deformation, and stress changes due to fluid diffusion. These complexities underscore the need for models that capture the diversity of fault slip behaviors observed in experimental settings.

%%%%%%%%%%%%%%%%%%%%%%%%% To Work On 
A significant body of research based on numerical modeling has focused on simulating fluid-induced seismic and aseismic slip on faults governed by rate-and-state friction. Early studies primarily examined single planar faults with simplified fluid-fault coupling through the effective normal stress, either under the quasi-dynamic approximation \citep{mcclure2011,almakari2019,dublanchet2019} or by including fully-dynamic ruptures \citep{larochelle2021}. Subsequent work expanded this understanding by incorporating poroelastic effects, either in simplified representations  \citep{noda2022} or in more comprehensive forms \citep{norbeck2015,heimisson2022}. These studies demonstrated the critical influence of bulk poroelastic properties, hydraulic diffusivity, and friction evolution on rupture stability during fluid injection.

The impact of fluid pressure on complex fault geometries, including rough faults \citep{maurer2020} and fault networks \citep{dieterich2015}, has also been investigated. However, these studies often relied on simplified assumptions, such as neglecting fluid flow or hydromechanical coupling. 
Experimental evidence has shown that permeability is highly sensitive to shear slip and normal stress, undergoing continuous changes throughout the earthquake cycle \citep{barton1985,esaki1999,lee2002c,im2018,im2019}. In light of this, \cite{mcclure2011} developed a rate-and-state friction model incorporating permeability changes during earthquakes. Similarly, \cite{cappa2018,cappa2019} introduced a slip-dependent permeability model based on the cubic law to study the role of aseismic slip during fluid injection in a rate-and-state fault. However, their approaches were limited to modeling a single seismic event, as they neglected the healing of permeability. These efforts emphasized the importance of modeling fault permeability enhancement and dilatancy, particularly in low-permeability host rocks such as shale layers and geothermal reservoirs. More recent work \citep{zhu2020,ozawa2024,dunham2024} has addressed these limitations by incorporating both permeability changes and recovery mechanisms, enabling the modeling of permeability evolution throughout multiple earthquake cycles.

Motivated by natural cases, more realistic and complex fault network geometries need to be incorporated in induced seismicity modeling. Fault systems in nature can be difficult to image and exhibit complex activation mechanisms upon fluid injection. These factors make the evaluation, prediction, and control of injection-induced earthquakes extremely challenging. To investigate fluid injection in fault networks, Discrete Fault Network (DFN) models have been widely used to predict which faults might be reactivated, typically under the assumption that the surrounding rock matrix is impermeable and fluid migration is confined to fault surfaces \citep{hicks1996,kohl2007,maxwell2015,Karvounis2022,ciardo2023}. While DFN models offer valuable insights into potential fault reactivation, they often rely on the assumption of a constant friction coefficient and Mohr-Coulomb failure, which limits their ability to accurately capture the dynamics of slip evolution. \cite{im2024} advanced this approach by introducing a three-dimensional DFN rate-and-state model. However, their model employed an analytical solution for fluid diffusion, which restricts the representation of strong hydro-mechanical coupling, including critical effects such as permeability changes due to deformation.

%Recent advances have incorporated fault slip dynamics and rate-and-state friction, although computational challenges persist in  modeling both fluid diffusion and complex fault interactions over realistic timescales \citep{dieterich2015,larochelle2021,im2024}. 

%Similarly, \cite{heimisson2022} used a spectral boundary-integral method to model fault slip in poroelastic solids, incorporating rate-and-state friction, dilatancy, and compaction. Their work showed the critical role of bulk poroelastic properties and hydraulic diffusivity in determining rupture stability during fluid injection, providing a more nuanced  understanding of fluid-fault coupling than traditional DFN or poroelastic models.

% Furthermore, \cite{ciardo2023} explored fluid injection in a two-dimensional DFN, demonstrating how fluid-driven aseismic slip propagates in fractured rock masses. This investigation provides insights into the spatial and temporal evolution of aseismic slip, which is crucial for understanding and mitigating induced seismicity \citep{mcClure}.

Each of these modeling approaches has advanced our understanding of fluid-induced fault slip in different ways: (1) by coupling fault slip with bulk poroelastic deformation \citep{norbeck2015,heimisson2022,noda2022}, (2) by incorporating dynamic permeability changes and frictions laws \citep{mcclure2011, cappa2018, cappa2019, zhu2020, ozawa2024,dunham2024} and (3) by capturing the complexities of fracture networks and geometry \citep{hicks1996,kohl2007,maurer2020, ciardo2023, im2024}. However, each method has limitations due to either computational complexity, simplified representations or assumptions about fault permeability and connectivity. 

In this study, we build upon recent advances by developing an integrated, computationally efficient approach for modeling fluid-induced seismicity in complex fault network geometries and fluid injection scenarios. Our method enables the simulation of both aseismic slip events and dynamic earthquakes during fluid injection. Fluid diffusion is modeled along fault planes, with migration between interconnected faults. In Section 2, we introduce our methodology, which couples an H-matrix accelerated Boundary Element Method (BEM) for fault slip with a finite volume method (FVM) for fluid pressure diffusion. The approach not only captures fault reactivation but also tracks the evolution of fault slip, distinguishing  between seismic and aseismic behavior. This distinction is particularly significant, given growing evidence that aseismic slip can outpace fluid diffusion, and eventually trigger larger earthquakes at substantial distances from the injection point. Coupling two methods, usually one volumetric method with a BEM, allows to relax some assumptions of BEM, while keeping good numerical efficiency \citep{ma2019b,wang2024}.
In Section 3, we showcase the method's capabilities through two simulation examples that illustrate how fluid-faulting interactions control seismic activity in discrete fault networks. Our modeling results are consistent with observed behaviors of induced earthquake sequences, suggesting the key role of aseismic slip migration as a process controlling fluid-faulting interactions leading to seismicity. Using these examples, we discuss the implications of our results for understanding the interplay between fluid flow, aseismic slip and seismic activity in fault networks within low permeability formations. The insights gained from this work offer valuable guidance for mitigating the risks of induced seismicity in energy-related subsurface operations.

%This highlights the need for an integrated and computationally efficient approach that combines the strengths of these methods while addressing their weaknesses to achieve a precise modeling of fluid-induced seismicity with realistic fault network geometries and fluid injection scenarios. Therefore, it is essential to understand the relative contributions of fluid pressure diffusion, slow slip and stress perturbation in earthquake rupture during fluid injection in discrete fault network simulations. This is important for two reasons: (1) some models suggest that the expansion of the seismicity cloud is solely driven by the diffusion of elevated fluid pressure, which activates critically stressed faults \citep{shapiro1997,shapiro2002}, and (2) other models, which link the potential magnitude of induced earthquakes to injection parameters, assume that fault rupture is entirely seismic and remains confined within the fluid-pressurized zone \citep{mcGarr2014,shapiro2011}. Thus, these models do not consider the observed variety of slip modes and fault interactions.

%%%%%%%%%%%%%%%%%%%%%%%%%%%%%%%%%%%%%%%%%%%%%%%%%%%%%%%%%%%%%%%%%%%%%%%%%%%%%%%%%%%%%%%%%%%%%%%%%%%%%%%
% METHODOLOGY
%%%%%%%%%%%%%%%%%%%%%%%%%%%%%%%%%%%%%%%%%%%%%%%%%%%%%%%%%%%%%%%%%%%%%%%%%%%%%%%%%%%%%%%%%%%%%%%%%%%%%%%
\section{Method}

\subsection{Modeling fault slip coupled with fluid diffusion}

We present a 2D hydro-mechanical model that integrates slip on a network of faults, governed by rate-and-state friction, with fluid diffusion along the fault surfaces. This model is capable of capturing fault interactions within geometrically complex fault networks, and of simulating seismic and aseismic slip. The model can be readily extended to accommodate the non-linear evolution of fault permeability with ongoing slip. 
The main governing equations are described in this section, with the numerical methods used to solve them in the subsequent subsections.

The quasi-dynamic equilibrium of shear stress on the faults leads to the following equation that couples the mechanical and hydraulical parts of our model: 
\begin{equation}
\underbrace{\vphantom{\log \left(\frac{V}{V_0}\right)}\tau_{\text{el}}(\Delta u )+\tau_{\text{load}}-\frac{\mu}{2\beta}V}_{\text{Shear traction on the faults}} = \underbrace{\vphantom{\log \left(\frac{V}{V_0}\right)}(\sigma_{\text{el}}(\Delta u )+\sigma_{\text{load}}-p)}_{\text{Effective normal traction}}\;\times\underbrace{f(V,\theta)\vphantom{\frac{\mu}{2\beta}V}}_{\text{Rate-and-state friction}}  .
\end{equation}
On the left-hand side, $\tau_{\text{el}}(\Delta u )$ is the static elastic shear stress due to the slip $\Delta u$ on each fault, $\tau_{\text{load}}$ the external shear stress loading on the faults, $V$ the slip velocity, $\mu$ the shear modulus, and $\beta$ the S-wave velocity. 
We adopt the quasi-dynamic approximation \citep{rice1993,cochard1996}, in which the radiation damping term $\frac{\mu}{2\beta}V$ represents the instantaneous dynamic shear stress reduction due to slip. It partially accounts for the elasto-dynamic effects due to wave propagation. 
On the right-hand side, the effective normal stress on the faults involves the static elastic normal stress due to slip $\sigma_{\text{el}}(\Delta u )$, the external normal stress load $\sigma_{\text{load}}$, and the pore fluid pressure $p$. 

The friction coefficient is governed by rate-and-state friction:
\begin{equation}
f(V,\theta) = f_0+a\log \left(\frac{V}{V_0}\right)+b\log \left(\frac{\theta V_0}{D_c}\right).
\end{equation}
It depends on the slip velocity $V$ and on a state variable $\theta$, and involves four parameters: a direct effect coefficient $a$, an evolution effect coefficient $b$, and a reference friction coefficient $f_0$ at a reference slip velocity $V_0$ \citep{marone1998}.
The evolution of the state variable is governed by the so-called aging law (though the model can accommodate other evolution laws):
\begin{equation}
\frac{\mathrm{d}\theta }{\mathrm{d}t}= 1 - \frac{V \theta}{D_c},
\end{equation}
where $D_c$ is a critical slip distance. 

The space and time evolution of fluid pressure on the faults is governed by the following diffusion equation, along each fault and across fault crossings:
\begin{equation}
\frac{\partial p}{\partial t}(\mathbf{x}, t) =\overrightarrow{\nabla}\cdot \left[\alpha(\mathbf{x})\overrightarrow{\nabla}p(\mathbf{x},t)\right]+\frac{q(\mathbf{x},t)}{\phi(\beta_f+\beta_{\phi})},
\end{equation}
where $\overrightarrow{\nabla}\cdot$ is the divergence operator, $\overrightarrow{\nabla} p$ the pressure gradient, $\alpha(\mathbf{x})=\frac{k(\mathbf{x})}{\phi \eta(\beta_{\phi}+\beta_{f})} $ the diffusion coefficient, $k(\mathbf{x})$ the fault zone permeability, which may depend on the position along the fault, $\phi$ the porosity of the fault zone, $\eta$ the dynamic viscosity of the fluid, $\beta_{\phi}=\frac{1}{\phi}\frac{\partial \phi}{\partial P}$ the compressibility of the fault zone pores, $\beta_{f}=\frac{1}{\rho}\frac{\partial \rho}{\partial P}$ the compressibility of the fluid ($\rho$ is the fluid density), and $q(\mathbf{x},t)$ the injection rate.
At junctions between two faults, diffusion happens in all the four possible directions.

%%%%%%%%%%%%%%%%%%%%%%%%%%%%%%%%%%%%%%%%%%%%%%%%%%%%%%%%%%%%%%%%%%%%%%%%%%%%%%%%%%%%%%%%%%%%%%%%%%%%%%%
% MECHANICAL MODEL
%%%%%%%%%%%%%%%%%%%%%%%%%%%%%%%%%%%%%%%%%%%%%%%%%%%%%%%%%%%%%%%%%%%%%%%%%%%%%%%%%%%%%%%%%%%%%%%%%%%%%%%
\subsection{Accelerated boundary element method for elastic stress interactions}

The elastic shear stress $\tau_{\text{el}}(\mathbf{x})$ and normal stress $\sigma_{\text{el}}(\mathbf{x})$ at a given position $\mathbf{x}$ along a fault system due to a spatially distributed slip $\Delta u$ is calculated using boundary integral equations \citep{tada1997, romanet2020,romanet2024}.
For a non-planar fault, they are given by the following integrals along the fault network:
\begin{equation}
\begin{split}
\tau_{\text{el}}(\mathbf{x}) &= \frac{\mu}{2\pi(1-\nu)}\int_{\text{faults}} K^t_{\text{grad}}(\mathbf{x},\mathbf{y})  \frac{\partial}{\partial y^t}\Delta u(\mathbf{y}) \mathrm{d}l(\mathbf{y})\\
&+ \frac{\mu}{2\pi(1-\nu)}\int_{\text{faults}} K^t_{\text{curv}}(\mathbf{x},\mathbf{y})   \kappa^t(\mathbf{y}) \Delta u(\mathbf{y})  \mathrm{d}l(\mathbf{y}),
\end{split}
\end{equation}
\begin{equation}
\begin{split}
\sigma_{\text{el}}(\mathbf{x}) &=\frac{\mu}{2\pi(1-\nu)}\int_{\text{faults}} K^n_{\text{grad}}(\mathbf{x},\mathbf{y})  \frac{\partial}{\partial y^t}\Delta u(\mathbf{y}) \mathrm{d}l(\mathbf{y})\\
&+ \frac{\mu}{2\pi(1-\nu)}\int_{\text{faults}} K^n_{\text{curv}}(\mathbf{x},\mathbf{y})    \kappa^t(\mathbf{y}) \Delta u(\mathbf{y})  \mathrm{d}l(\mathbf{y}),
\end{split}
\end{equation}
where $\mu$ is the shear modulus and $\nu$ the Poisson's ratio of the host rock. The derivative of the slip along a fault is given by:
\begin{equation}
\frac{\partial}{\partial y^t}\Delta u(\mathbf{y}) = \left(n_2(\mathbf{y})\frac{\partial}{\partial y_1}-n_1(\mathbf{y})\frac{\partial}{\partial y_2}\right)\Delta u(\mathbf{y}),
\end{equation}
where $n_1$ and $n_2$ are respectively the first and second component of the local normal vector to the fault $\mathbf{n}(\mathbf{y})$.
The absolute value of the local curvature along the fault $|\kappa^t(\mathbf{y})|$ can be interpreted geometrically as the inverse of the radius of the tangential circle to the fault. The stress transfer kernels $K^t_{\text{grad}}(\mathbf{x},\mathbf{y})$, $K^t_{\text{curv}}(\mathbf{x},\mathbf{y})$, $K^n_{\text{grad}}(\mathbf{x},\mathbf{y})$ and $K^n_{\text{curv}}(\mathbf{x},\mathbf{y})$ depend only on the fault network geometry, and are given in appendix \ref{sec:kernels} (see also eqs. C7 and C8 in \cite{romanet2020}).

The two above equations are discretised by decomposing the fault network geometry into a grid of non-overlapping linear straight elements and by assuming constant slip within each element \citep{romanet2020}. A constraint on the fault grid generation is that fault crossings must lie at the edge of neighboring elements, as shown in Fig.~\ref{fig:fault_crossing}. Stresses are evaluated at the center of each element. After discretization, stresses are computed from slip values through a matrix-vector product between the boundary element matrix and the vector of slip values at all elements. This operation can be computationally expensive due to the fully-populated nature of the matrix. To speed-up the stress evaluation, a data-sparse representation of the matrix is obtained with the hierarchical matrix technique \citep{hackbusch1999, borm2003, chaillat2017} (see \cite{ohtani2011,bradley2014,sato2021,ozawa2023} for previous applications in  earthquake modeling studies).

%%%%%%%%%%%%%%%%%%%%%%%%%%%%%%%%%%%%%%%%%%%%%%%%%%%%%%%%%%%%%%%%%%%%%%%%%%%%%%%%%%%%%%%%%%%%%%%%%%%%%%%
% HYDRAULIC MODEL
%%%%%%%%%%%%%%%%%%%%%%%%%%%%%%%%%%%%%%%%%%%%%%%%%%%%%%%%%%%%%%%%%%%%%%%%%%%%%%%%%%%%%%%%%%%%%%%%%%%%%%%
\subsection{Finite volume method for fluid diffusion along faults}
%The diffusion equation is given by:
%\begin{equation}
%\frac{\partial p}{\partial t}(\mathbf{x},t) =\overrightarrow{\nabla}\cdot \left[ \alpha(\mathbf{x})\overrightarrow{\nabla}p(\mathbf{x},t)\right]+\frac{q(\mathbf{x},t)}{\phi(\beta_f+\beta_{\phi})},
%\end{equation}

The classic finite volume method (FVM) decomposes the fluid diffusion domain into control volumes $\Gamma_i$.
In our case, considering diffusion along faults, each ``volume'' $\Gamma_i$ is a linear sub-segment of a fault, with length $L_i$.
The primary variables in FVM are the averages of pressure $p$ and injection rate $q$ over each control volume, defined by
\begin{equation}
\begin{split}
P_i(t) &=\frac{1}{L_i} \int_{L_i} p(\mathbf{x},t)\mathrm{d}l(\mathbf{x}),\\
Q_i(t) &=\frac{1}{L_i} \int_{L_i} q(\mathbf{x},t)\mathrm{d}l(\mathbf{x}).\\
\end{split}
\end{equation}
% \begin{equation}
%  \frac{\partial  P_i}{\partial t}(t)  = \frac{1}{L_i} \sum_{i} \int_{L_i} \alpha(\mathbf{x}) \overrightarrow{\nabla}P(\mathbf{x},t) dS(\mathbf{x}) + \frac{Q_i(t)}{\phi(\beta_f+\beta_{\phi})}
% \end{equation}
These averages are governed by the following spatially-discretized equations \citep{patankar2018}:
\begin{equation}
\begin{split}
 \frac{\partial P_i  }{\partial t}(t) &=  \alpha_{i,i+1} \frac{P_{i+1}(t) -P_{i}(t) }{\Delta x_{i,i+1}^2}-\alpha_{i-1,i}\frac{P_{i}(t) -P_{i-1}(t) }{\Delta x_{i-1,i}^2}+\frac{Q_i(t)}{\phi(\beta_f+\beta_{\phi})},
 \end{split}
\end{equation}
where $\alpha_{i,i+1}=\frac{2\Delta x_{i,i+1}\alpha_{i}\alpha_{i+1}}{L_{i+1}\alpha_i+L_{i}\alpha_{i+1}}$ is the harmonic average of the diffusivity between two neighboring elements $i$ and $i+1$, and $\Delta x_{i,i+1}=\frac{L_i+L_{i+1}}{2}$ is the distance between the center of elements $i$ and $i+1$.

The temporal discretization of the governing equations can be done by either explicit or implicit time schemes.
Using an explicit scheme, which is conditionally stable, is not efficient because of the small time step required to ensure numerical stability. To circumvent this limitation, we adopt an implicit time scheme, which is unconditionally stable: 
\begin{equation}
\begin{split}
\frac{ P_{i}^{k+1}- P_{i}^{k}}{\Delta t}  &=   \alpha_{i,i+1} \frac{P_{i+1}^{k+1}-P_{i}^{k+1}}{\Delta x_{i,i+1}^2}- \alpha_{i-1,i}\frac{P_{i}^{k+1}-P_{i-1}^{k+1}}{\Delta x_{i-1,i}^2}+  \frac{Q_i^{k+1}}{\phi(\beta_f+\beta_{\phi})},
 \end{split}
\end{equation}
 where the superscript $k+1$ stands for time $t_{k+1}=t_k+\Delta t$, where $\Delta t$ is the time step.  The discrete equation can be written in the form of a matrix-vector multiplication:
\begin{equation}
\mathbf{A}\mathbf{P}^{k+1} = F(\mathbf{P}^k,\mathbf{Q}^{k+1})
\label{FVM}
\end{equation}
where $\mathbf{P}^{k}$ is the vector of pressure values at time $t_{k}$, $\mathbf{Q}^{k+1}$ is the injection rate at the new time $t_{k+1}$, $F$ is a linear function and $\mathbf{A}$ a tri-diagonal matrix that depends on $\Delta t$ and constant parameters. This system of linear equations has to be solved for $\mathbf{P}^{k+1}$ to update the fluid pressure at the new time $t_{k+1}$. 
In this work, it is solved with a conjugate gradient method.

\begin{figure}[H]% discretized fault
\centering
\includegraphics[width=\textwidth]{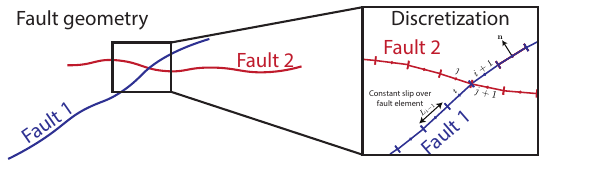}
\caption{The numerical grid for both Boundary Element Method (BEM) and Finite Volume Method (FVM). The connection between two faults is done at the element edges.}
\label{fig:fault_crossing}
\end{figure}

The classic implicit FVM described above is for a single fault. We introduce a modification for two crossing faults, as required to handle diffusion along a fault network.
At the junction between two faults, diffusion happens in all the four possible directions. We further require fault junctions to lie at the edge of neighboring control volumes, as shown in Fig. \ref{fig:fault_crossing}. 
Let us consider the junction between faults 1 and 2, lying at the edge between elements $i$ and $i+1$ on fault 1 and between elements $j$ and $j+1$ on fault 2.  
To account for the effect of diffusion along fault 2, the FVM scheme for element $i$ of fault 1 is updated to 
\begin{equation}
\begin{split}
  \frac{ P_{i}^{k+1}-P_i^k}{\Delta t}   &= \underbrace{  \alpha_{i,i-1} \frac{P_{i-1}^{k+1}-P_{i}^{k+1}}{(\Delta x_{i-1,i})^2}}_{\text{Effect of the left element}}+\underbrace{\alpha_{i,i+1}  \frac{P_{i+1}^{k+1}-P_{i}^{k+1}}{(\Delta x_{i,i+1})^2}}_{\text{Effect of the right element}}+\underbrace{ \vphantom{\frac{1}{\mu}} \frac{Q_i^{k+1}}{\phi (\beta_{\phi}+\beta_f)}}_{\text{Source term}} \\
 &+  \underbrace{  \alpha_{i,j} \frac{P_{j}^{k+1}-P_{i}^{k+1}}{(\Delta x_{i,j})^2}}_{\text{Effect of fault 2, element j }}+  \underbrace{  \alpha_{i,j+1} \frac{P_{j+1}^{k+1}-P_{i}^{k+1}}{(\Delta x_{i,j+1})^2}}_{\text{Effect of fault 2,  element j+1}}.
 \end{split}
 \label{eq:FVM2}
\end{equation}
The first line in eq. \eqref{eq:FVM2} is the same as in a single fault. The second line accounts for the coupling of fluid pressure between the two faults. The other three elements attached to the fault crossing are updated in a similar way.
With this modification, the matrix $\mathbf{A}$ is no longer tri-diagonal.

% \begin{equation}
% \begin{split}
% P_{i}^{k+1}   &= \underbrace{  \frac{\Delta t k_{i,i-1}}{\mu  \phi (c_r+c_f)} \frac{P_{i-1}^{k+1}-P_{i}^{k+1}}{(\Delta x)^2}}_{\text{Effect of the left element}}+\underbrace{ \frac{\Delta t k_{i,i+1}}{\mu  \phi (c_r+c_f)}  \frac{P_{i+1}^{k+1}-P_{i}^{k+1}}{(\Delta x)^2}}_{\text{Effect of the right element}}+\underbrace{ \vphantom{\frac{1}{\mu}} \frac{q_i\Delta t}{\rho  \phi (c_r+c_f)}}_{\text{Source term}} +P_i^k \\
%  &+  \underbrace{  \frac{\Delta t k_{i,j}}{\mu  \phi (c_r+c_f)} \frac{P_{j}^{k+1}-P_{i}^{k+1}}{(\Delta x)^2}}_{\text{Effect of fault 2, j element}}+  \underbrace{  \frac{ \Delta t k_{i,j+1}}{\mu  \phi (c_r+c_f)} \frac{P_{j+1}^{k+1}-P_{i}^{k+1}}{(\Delta x)^2}}_{\text{Effect of fault 2, j+1 element}}+
%  \end{split}
% \end{equation}

% \begin{equation}
% \begin{split}
% &\left(1+ \frac{\Delta t (k_{i,i-1}+k_{i,i+1})}{\mu  \phi (c_r+c_f)(\Delta x)^2} +\frac{\Delta t (k_{i,j}+k_{i,j+1})}{\mu  \phi (c_r+c_f)(\Delta x)^2}  \right)P_{i}^{k+1}  \\
% &- \frac{\Delta t k_{i,i-1}}{\mu  \phi (c_r+c_f)} \frac{P_{i-1}^{k+1}}{(\Delta x)^2}\\
% &- \frac{\Delta t k_{i,i+1}}{\mu  \phi (c_r+c_f)} \frac{P_{i+1}^{k+1}}{(\Delta x)^2} \\
% &- \frac{\Delta t k_{i,j}}{\mu  \phi (c_r+c_f)} \frac{P_{j}^{k+1}}{(\Delta x)^2} \\
% &- \frac{\Delta t k_{i,j+1}}{\mu  \phi (c_r+c_f)} \frac{P_{j+1}^{k+1}}{(\Delta x)^2}\\
%  &= \frac{q_i\Delta t}{\rho  \phi (c_r+c_f)} +P_i^k 
%  \end{split}
% \end{equation}

\subsection{Adaptive-time solver for the hydro-mechanical model}

We adopt the same fault network discretisation for the BEM and FVM.
We couple the FVM diffusion solver and the mechanical ODE solver (Bulirsch–Stoer) as follows: 

\begin{enumerate}
    \item An intended time step $\Delta t_{\text{try}}$ is suggested by the adaptive ODE solver.
    \item The pressure $P^{k+1}$ at time $t_{k+1}=t_k+\Delta t_{\text{try}}$ is updated using the FVM (equation~\ref{FVM}). 
    \item Assuming the pressure is constant and equal to $P^{k+1}$ within the current time interval $[t_k,t_{k+1}]$, the ODE solver updates the slip velocity $V^{k+1}$ and the state variable $\theta^{k+1}$. 
    \item If Step 3 is successful, the update at time $t_{k+1}$ is now complete and we proceed to the next time $t_{k+2}$ from Step 1. Otherwise, the update has failed and we return to Step 1 to suggest a new time step. 
\end{enumerate}

\section{Examples of slip behavior and seismicity induced by fluid injection in fault networks}

To showcase the capabilities of this seismo-hydro-mechanical model, we simulate two conceptual examples of fluid injection, each representative of realistic field-scale scenarios inspired by fault activation experiments and reservoir stimulation processes. The first example examines aseismic slip induced by fluid injection on a rate-strengthening fault, which subsequently triggers microseismicity on nearby secondary faults. The second example explores fluid diffusion and induced slip within an interconnected fault network. In both scenarios, we adopt a simplified representation of hydraulic properties, assuming constant fault permeability (i.e., no deformation-permeability coupling) and an impermeable surrounding rock (i.e., neglecting poroelastic effects). The injected fluid is water at constant temperature.
%%%%%%%%%%%%%%%%%%%%%%%%%%%%%%%%%%%%%%%%%%%%%%%%%%%%%%%%%%%%%%%%%%%%%%%%%%%%%%%%%%%%%%%%%%%%%%%%%%%%%%%
% APPLICATION TO MICRO-SEISMICITY
%%%%%%%%%%%%%%%%%%%%%%%%%%%%%%%%%%%%%%%%%%%%%%%%%%%%%%%%%%%%%%%%%%%%%%%%%%%%%%%%%%%%%%%%%%%%%%%%%%%%%%%
\subsection{Microseismicity triggered near a rate-strengthening fault stimulated by fluid injection}
\label{microseismicity}

We simulate localized fluid injection at the center of a 10-m-long permeable fault, surrounded by a network of smaller, disconnected and randomly oriented secondary faults. The objective is to demonstrate how fluid injection induces aseismic slip on a rate-strengthening fault, which subsequently triggers microseismic activity on adjacent secondary rate-weakening faults through elastic stress transfer. Values of model parameters are given in Tab. \ref{tab:microseismicity}.
% Geometry
The 2D model spans 15 m by 8 m. The main fault is represented as a self-similar geometrically rough fault \citep{power1991,renard2006,candela2011,dunham2011b}, with deviations from planarity characterized by an amplitude-to-wavelength ratio of $5\times 10^{-4}$. Surrounding this pressurized fault are 800 secondary faults with equal length, each $100$ times shorter than the main fault. These secondary faults are oriented with variable angles, randomly drawn from a uniform distribution between $-30^{\circ}$ and $+30^{\circ}$ relative to the main fault's orientation. 
% Fault properties
Before injection, the hydraulic and frictional properties of the faults, as well as the elastic properties of the surrounding rocks, are assumed to be uniform and set to typical values consistent with crustal reservoir conditions (Table 1). The pressurized main fault is rate-strengthening ($a>b$) and thus can slip aseismically. The secondary faults are velocity-weakening ($a<b$) and thus can rupture dynamically when the Coulomb stress changes caused by slip on the main fault are positive and sufficiently large. To ensure computational efficiency, each secondary fault is modeled with a single element, allowing us to capture the temporal evolution of their moment release while omitting the spatial details of their rupture propagation.

We consider a faulted crustal reservoir in an intraplate tectonic regime, where external stressing rate is very low or inexistent and the system is critically stressed (e.g. \cite{townend2000}). To replicate the natural boundary conditions observed in most intraplate regions where induced seismicity occurs, we do not impose any tectonic loading \citep{larochelle2021}. This approach is justified by the fact that stress changes due to fluid injection typically occur much more rapidly than those caused by tectonic loading. This is evidenced by the sharp increase in seismicity rates in the central US following the onset of wastewater injection \citep{ellsworth2013,magnani2017}. In this context, the fault slip history in our model is governed solely by the stress perturbations resulting from fluid injection and diffusion. 

At the start of the simulation, the system is allowed to evolve for 10 years to minimize the influence of initial conditions on the subsequent injection-induced reactivation. The initial conditions are such that all faults are initially sliding at steady state (i.e. $V_0\theta_0/D_c=1$). Following this period of model initialization, fluid is injected at a constant rate ($q = 1.3 \times 10^{-6}\;\mathrm{m \;s}^{-1}$) at a grid point on the main fault for 5,000 seconds, reaching a peak pressure of 3 MPa. In the simulation, the evolution of fluid pressure, fault slip, slip velocity, stress and friction are calculated. Slip is defined to be seismic when slip velocity exceeds a threshold of $10^{-3} \; \mathrm{m \;s}^{-1}$.

\begin{figure}[H]% discretized fault
\centering
\includegraphics[width=\textwidth]{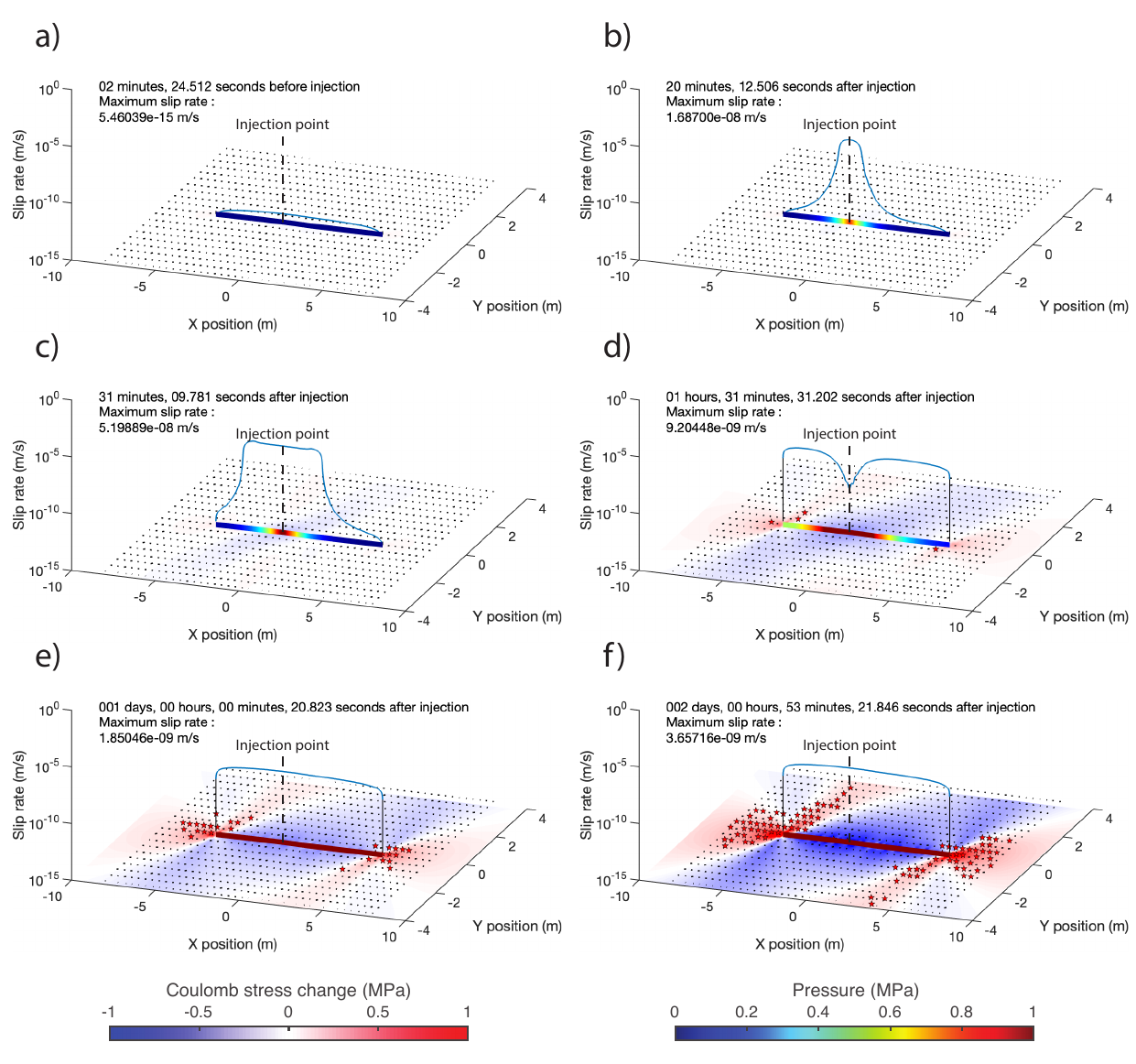}
\caption{Evolution of injection-induced slow slip on a main fault and triggered earthquakes on secondary faults.
In each panel, the geometry of the fault network (comprising one main fault and multiple secondary faults) is shown at the bottom, the spatial distribution of fluid pressure is shown along the main fault (color scale on the bottom right), slip velocity profiles are shown as curves above the main fault (in logarithmic scale, vertical axis),
secondary faults that have slept seismically are indicated by red stars,
and the 2D spatial distribution of Coulomb stress along slip planes parallel to the main fault is shown in colors at the bottom (color scale on the bottom left).
The selected snapshot times for each panel, relative to the onset of injection, are:
(a) $-2$ minutes (right before injection), (b) $20$ minutes, (c) $31$ minutes, (d) $1$ hour $31$ minutes ($8$ minutes after the end of injection), (e) $1$ day and (f) $2$ days.}
\label{fig:triggered_seismicity}
\end{figure}

Figure \ref{fig:triggered_seismicity} illustrates the various stages of injection-induced activity. Before injection begins, the fault is sliding very slowly, with a maximum slip rate of $5.4 \times 10^{-15} \; \mathrm{~m \;s}^{-1}$ (Fig. \ref{fig:triggered_seismicity}.a). Once fluid injection starts, a slow aseismic slip transient begins to propagate from the injection point (Fig. \ref{fig:triggered_seismicity}.b). As injection continues and fluid diffuses along the fault, this slow aseismic slip propagates further (Fig. \ref{fig:triggered_seismicity}.c). Interestingly, the slip front and the fluid pressure front do not propagate synchronously. Instead, the slow slip outpaces fluid diffusion. 
Eight minutes after injection ceases, the slow slip front has already reached the edges of the fault, whereas fluid pressure continues to diffuse within the fault (Fig. \ref{fig:triggered_seismicity}.d). 
Theoretical models of single ``critically stressed" faults—whether governed by constant friction \citep{bhattacharya2019,viesca2021} or by slip-weakening friction \citep{garagash2013,saez2022}— similarly predict that the slow slip front can outpace the pressure diffusion front. In our case, we observe a similar result on a rate-and-state fault. 

The slip on the main fault transfers stress to nearby smaller faults, reactivating them and triggering micro-seismicity. 
The onset of seismicity corresponds to the regions of positive Coulomb stress changes concentrated at the edges of the main fault during the slow slip phase. Long after the end of injection (Figs. \ref{fig:triggered_seismicity}.e and f respectively), even though the fluid pressure front has fully diffused across the fault, seismicity continues to propagate outward from the fault tips, driven by the accumulation of positive Coulomb stress changes. 

\begin{figure}[H]% discretized fault
\centering
\includegraphics[width=\textwidth]{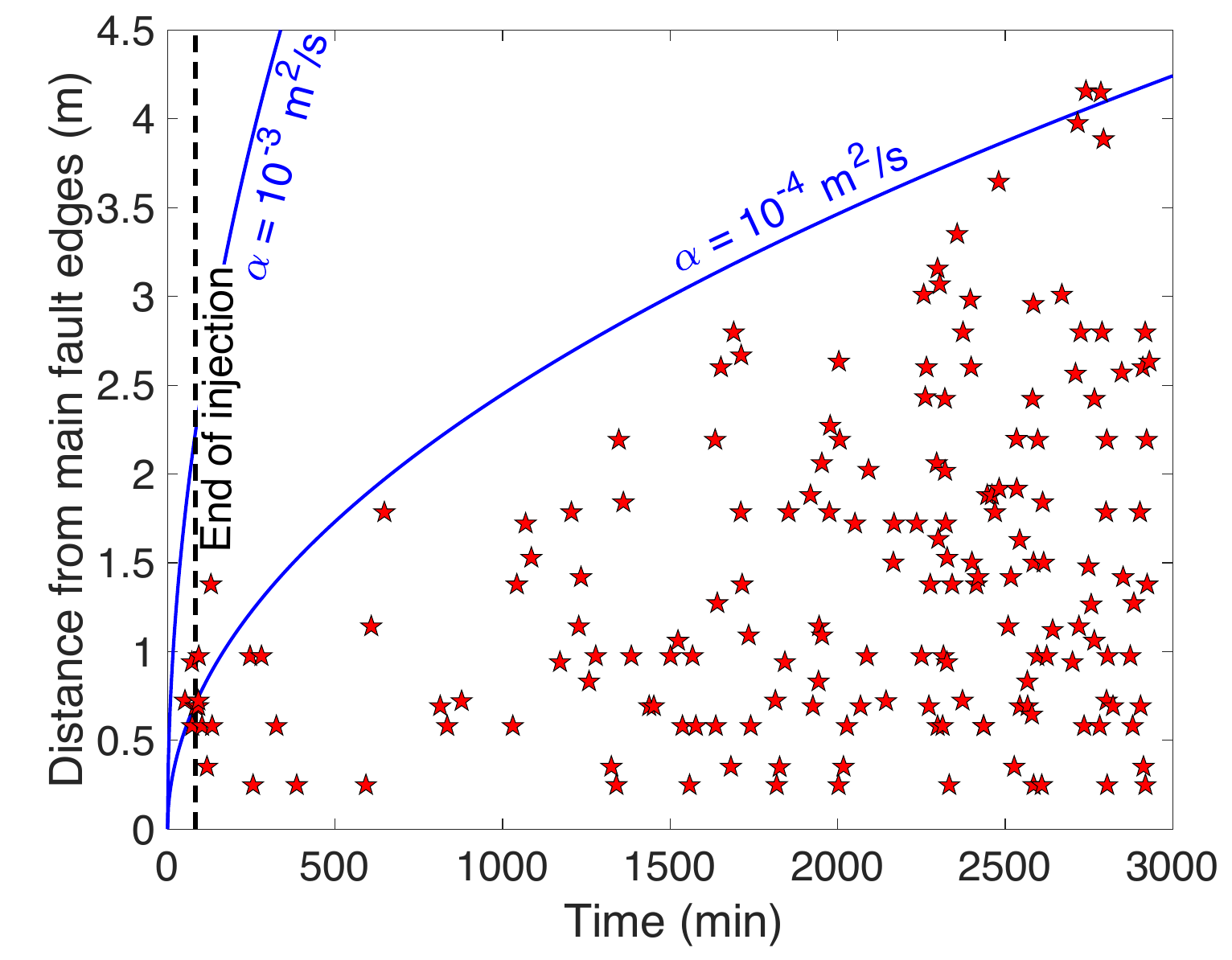}
\caption{Space-time distribution of seismicity. Time is relative to the start of injection and distance is relative the closest main fault tip. Blue curves are examples of apparent diffusion fronts, with values of effective diffusivity $\alpha$ indicated in labels. Although fluid diffusion is confined within the main fault, the seismicity front outside the main fault exhibits a diffusion-like migration pattern. 
}
\label{fig:diffusion_microseismicity}
\end{figure}

To better understand the evolution of microseismicity, we analyzed its time-dependent migration away from the nearest fault tip, which persists for over two days after the end of injection (Fig. \ref{fig:diffusion_microseismicity}). This is particularly important as field observations have shown that injection-induced aseismic deformation can alter local stress fields, controlling the migration of seismicity and, in some cases, reactivating  misoriented faults \citep{schoenball2014, duboeuf2021, goebel2017}. 
%However, in natural-scale cases it is challenging to decouple the effect of elastic stress state transfer and the poroelastic component due to fluid diffusion in the bulk associated to permeability enhancement in the vicinity of the fault. 
In our model, the seismicity patterns are caused exclusively by elastic stress transfer from the slow-slip event. %the Coulomb stress change due to the propagation of the slow slip front. 
The spatio-temporal distribution of seismicity (Fig. \ref{fig:diffusion_microseismicity}) exhibits an apparent diffusion-like pattern, where the distance of the seismicity front scales with the square root of time, consistent with \citep{shapiro2002}. For a diffusion process, the general scaling relation is: 
\begin{equation}
x = \sqrt{\alpha t},
\label{eq:difusion}
\end{equation}
where $x$ represents the distance from the source of the perturbation (i.e., injection point), in our case the tip of the fault, and $t$ is the elapsed time. In our simulation, we imposed a constant fluid diffusivity in the main fault of  $\alpha= 1\times 10^{-3}\;\mathrm{m}^2\;\mathrm{s}^{-1}$ and found that the fluid diffusion occurs an order of magnitude faster than the microseismicity diffusion, which has an effective diffusivity of $\alpha= 1\times 10^{-4}\;\mathrm{m}^2 \;\mathrm{s}^{-1}$. 
This result suggests that estimates of hydraulic diffusivity based solely on the migration of the microseismicity front, as proposed by \cite{shapiro1997}, may not be accurate. Our finding is consistent with previous modeling works on fluid injection into permeable, deformable faults, suggesting that the seismic migration depends on the fault properties and stress criticality that drive the aseismic slip rather than on the hydraulic diffusivity \citep{wynantsMorel2020,deBarros2021}. 

\begin{table}[H]
\scalebox{0.8}{
\begin{tabular}{cll}
  \hline
 \multicolumn{3}{c}{Frictional properties of the main fault} \\
  \hline
$a$&0.006 & Direct effect\\
$b$&0.005 & Evolution effect\\
$D_c$&$1\; \mathrm{\mu m}$ & Critical Distance\\
$f_0$&0.6 & Reference friction\\
$V_0$&$1\times10^{-6}\;\mathrm{m \;s}^{-1}$ & Reference velocity\\
 \hline
 \multicolumn{3}{c}{Frictional properties for small faults} \\
  \hline
$a$& 0.005& Direct effect\\
$b$& 0.007& Evolution effect\\
$D_c$&  $4\times10^{-3}\; \mathrm{\mu m}$& Critical Distance\\
$f_0$&0.6 & Reference friction\\
$V_0$&$1\times10^{-6}\;\mathrm{m\;s}^{-1}$ & Reference velocity\\
\hline
\multicolumn{3}{c}{Rock elastic properties} \\
\hline
$\mu$&$3.2040\times 10^{10}\;\mathrm{Pa}$ & Shear modulus\\ 
$\nu$& 0.1541& Poisson's ratio \\
\hline
\multicolumn{3}{c}{Fault hydraulic and fluid properties} \\
\hline
$k$&$1\times 10^{-15}\;\mathrm{m}^2$ & Permeability\\
$\beta_f$&$ 9\times 10^{-9}\;\mathrm{Pa}^{-1}$& Compressibility of the fluid \\
$\beta_{\phi}$&$1\times 10^{-9}\;\mathrm{Pa}^{-1}$ & Compressibility of the pore\\
$\phi$& 0.1& Porosity \\  
$\eta$& $1\times 10^{-3}\;\mathrm{Pa\;s}$& Dynamic viscosity \\
$q$&$1.3416\times10^{-6} \;\mathrm{m\;s}^{-1}$ & Flow rate of injection \\
$t_{\text{inj}}$& $10 \text{ years}$& Beginning of injection \\
$\Delta t_{\text{inj}}$& $5000\;\mathrm{s}$ & Duration of injection \\
\hline
\multicolumn{3}{c}{Initial values for mechanical conditions} \\
\hline
$V_{\text{ini}}$& $10^{-12}\;\mathrm{m\;s}^{-1}$ & Initial slip velocity\\
$\theta_{\text{ini}}$&$10^{6}\;\mathrm{s}$ & Initial state\\
$\sigma_{\text{load}}$  & $5\;\mathrm{MPa} $& Normal stress\\
\hline
\multicolumn{3}{c}{Length scales} \\
\hline
$ds$&$5\;\mathrm{mm}$ & Grid size \\  
$L_b$&$1.3329\;\mathrm{m}$ & Process zone size
\end{tabular}}
\caption{Parameter values for the simulation of micro-seismicity triggered by fluid injection. $L_b$ is the process zone size given by \cite{rubin2005}.}
\label{tab:microseismicity}
\end{table}

%%%%%%%%%%%%%%%%%%%%%%%%%%%%%%%%%%%%%%%%%%%%%%%%%%%%%%%%%%%%%%%%%%%%%%%%%%%%%%%%%%%%%%%%%%%%%%%%%%%%%%%
% APPLICATION TO DIFFUSION ALONG FAULT NETWORK
%%%%%%%%%%%%%%%%%%%%%%%%%%%%%%%%%%%%%%%%%%%%%%%%%%%%%%%%%%%%%%%%%%%%%%%%%%%%%%%%%%%%%%%%%%%%%%%%%%%%%%%

\subsection{Fluid diffusion and aseismic slip propagation
in a fault network}
\label{network}

\begin{figure}[H]% discretized fault
\centering
\includegraphics[width=0.5\textwidth]{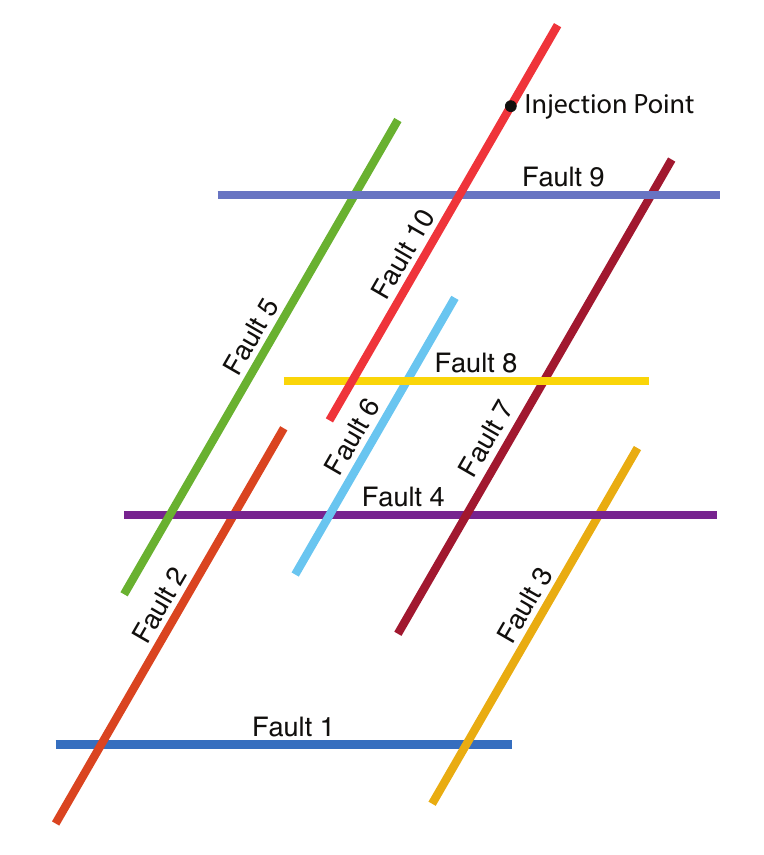}
\caption{Geometry of the fault network.}
\label{fig:network_geometry}
\end{figure}

In the second example, we examine the response of a hydraulically connected network of faults to fluid injection. The geometry of the fault network is defined in Fig.~\ref{fig:network_geometry} and parameter values are listed in Tab.~\ref{tab:table_network}. Unlike the first example, all the faults in this scenario are planar, have comparable lengths, and the rupture on each fault is spatially well-resolved with grid sizes smaller than the process zone size \citep{rubin2005}. The frictional properties are uniform across all faults, exhibiting a rate-strengthening behavior. Slip occurs in mode II (in plane shear), with interactions between faults facilitated by static stress transfer. The injection procedure involves constant injection rate at $2\times 10^{-6}\; \mathrm{m\;s}^{-1}$ over a $2000$ seconds time interval, pressurizing the fluid up to $1.6$ MPa. As injection begins, fluid diffusion occurs throughout the fault network.

%%%%%%%%%%%%%%%%%%%%%%%%%%%%%%%%%%%%%%%%%%%%%%%%%%%%%%%%%%%%%%%%%%%%%%%%%%%%%%%%%%%%%
% FROM HERE 
%%%%%%%%%%%%%%%%%%%%%%%%%%%%%%%%%%%%%%%%%%%%%%%%%%%%%%%%%%%%%%%%%%%%%%%%%%%%%%%%%%%%%

\begin{figure}[H]% discretized fault
\centering
\includegraphics[width=0.60\textwidth]{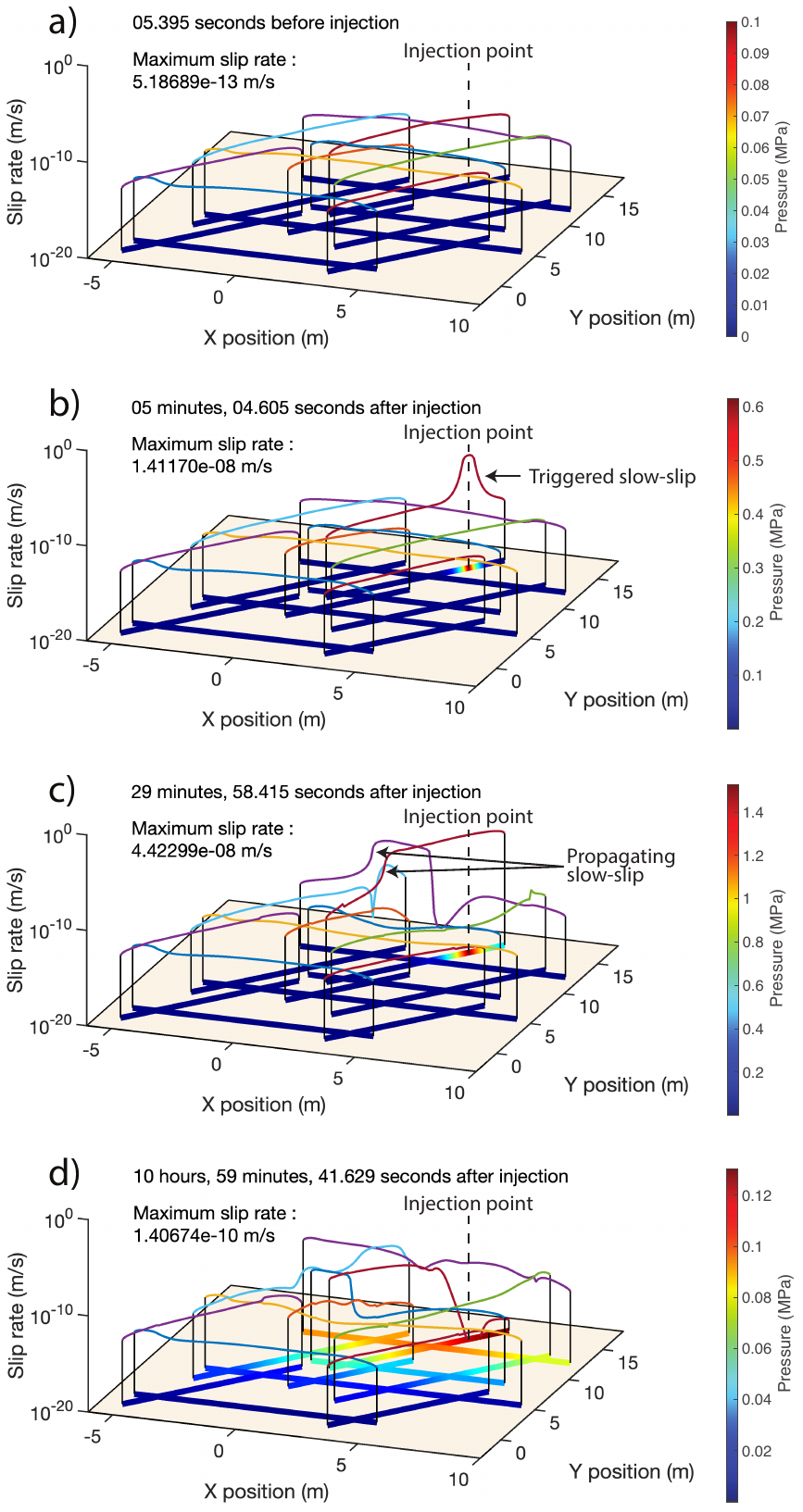}
\caption{Snapshots of slip velocity and fluid pressure along the fault network. Selected times, relative to the onset of injection, are: (a) $-5$~seconds (right before injection starts), (b) 5 minutes, (c) 30 minutes, (d) 11 hours (10.5 hours after injection stops). In each panel, the fault network geometry is shown in 2D view at the bottom, together with the fluid pressure (color scale on the right side). Slip velocity profiles are shown as curves above each fault, in logarithmic scale (vertical axis).}
\label{fig:snapshot_network}
\end{figure}

\begin{figure}[H]% discretized fault
\centering
\includegraphics[width=0.5\textwidth]{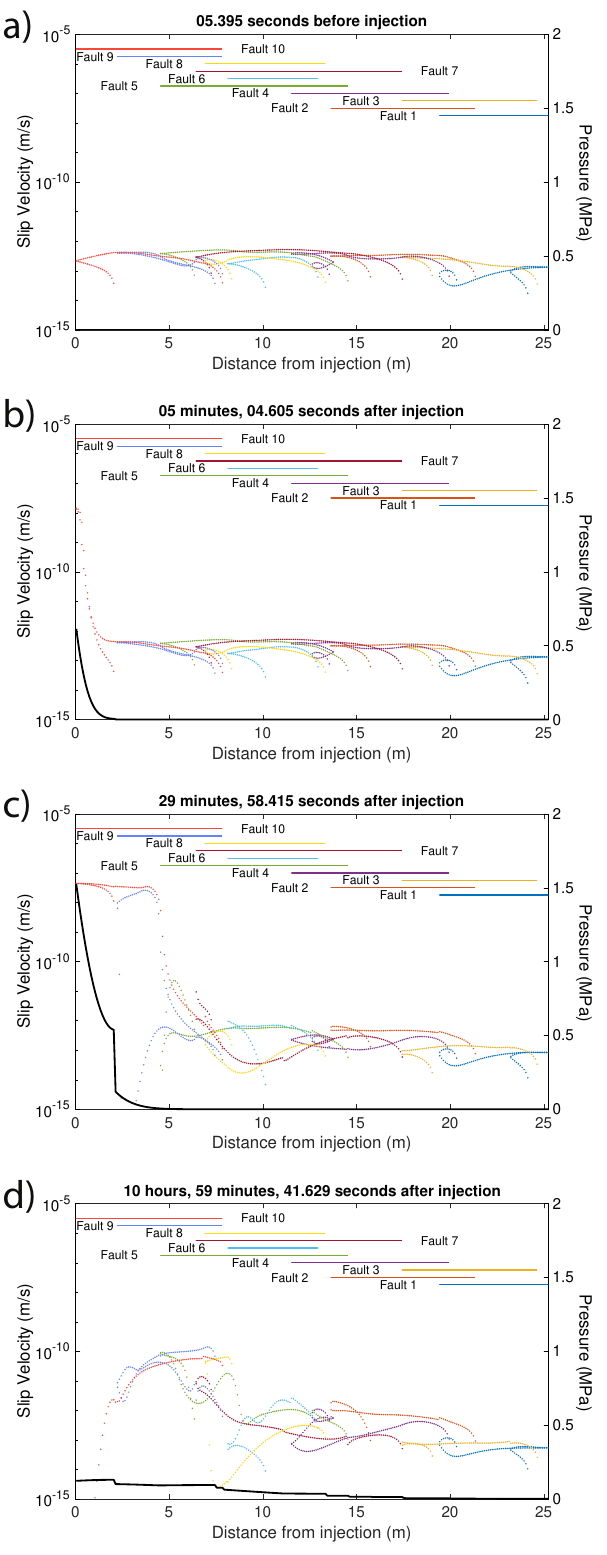}
\caption{Snapshots of slip velocity and fluid pressure as a function of distance from the injection point. The snapshot times are the same as in Fig.~\ref{fig:snapshot_network}. In each panel, the extent of each fault is indicated by colored horizontal lines at the top, and their slip velocities by curves of the corresponding color. The black curve shows the maximum pressure across the fault network at each distance from the injection point.  
}
\label{fig:network_dist}
\end{figure}

The response of the fault network to fluid injection is visualized through the snapshots of slip velocity and fluid pressure shown in 2D view in Fig. \ref{fig:snapshot_network} 
and as a function of distance from the injection point in Fig. \ref{fig:network_dist}. 
Before injection, all faults are slowly slipping due to a small applied shear load of $0.001\; \mathrm{Pa.s}^{-1}$ (Figs. \ref{fig:snapshot_network}.a and \ref{fig:network_dist}.a). 
% Just after injection
Injection is then prescribed on Fault 10. Shortly after, fluid pressure at the injection point increases rapidly, triggering a slow slip event along the injection fault within 5 minutes (Fig.~\ref{fig:snapshot_network}.b). In these early stages, there is an approximately linear relation between the rise in fluid pressure and the logarithm of slip velocity on Fault 10 (Fig.~\ref{fig:network_dist}.b), as expected in rate-and-state models when the changes in slip and state variable are still small \citep{dublanchet2019,larochelle2021}.
% End of injection
As the injection ends, 30 minutes after it started, a complex slip history develops across the fault network (Fig. \ref{fig:snapshot_network}.c and \ref{fig:network_dist}.c). Notably, while fluid diffusion remains largely confined within the injection fault (Fault 10), the slow slip propagates ahead of the pressure front, reaching Faults 9 and 5 (Figs.~\ref{fig:snapshot_network}.c). From Fig.~\ref{fig:network_dist}.c we can discriminate that the slow slip on Fault 5 is driven by the redistribution of stress across faults rather than by the fluid diffusion process. Indeed, pressure suddenly drops well before reaching Fault 5, where Fault 10 intersects Fault 9, as fluid diffusion transitions from a single pathway to three directions.
% Long after injection
Well after injection has ended, after 11 hours, the fluid pressure has decreased to minimum values and its front is slowly diffusing along the fault network (Fig. \ref{fig:snapshot_network}.d and \ref{fig:network_dist}.d). Nonetheless, the fault network is still quite active: faults far from the injection point (ie. Faults 5, 7 and 8) are still slowly slipping while the injection fault (Fault 10) and its first connected fault (Fault 9) are decelerating. 

\begin{figure}[H]% discretized fault
\centering
\includegraphics[width=\textwidth]{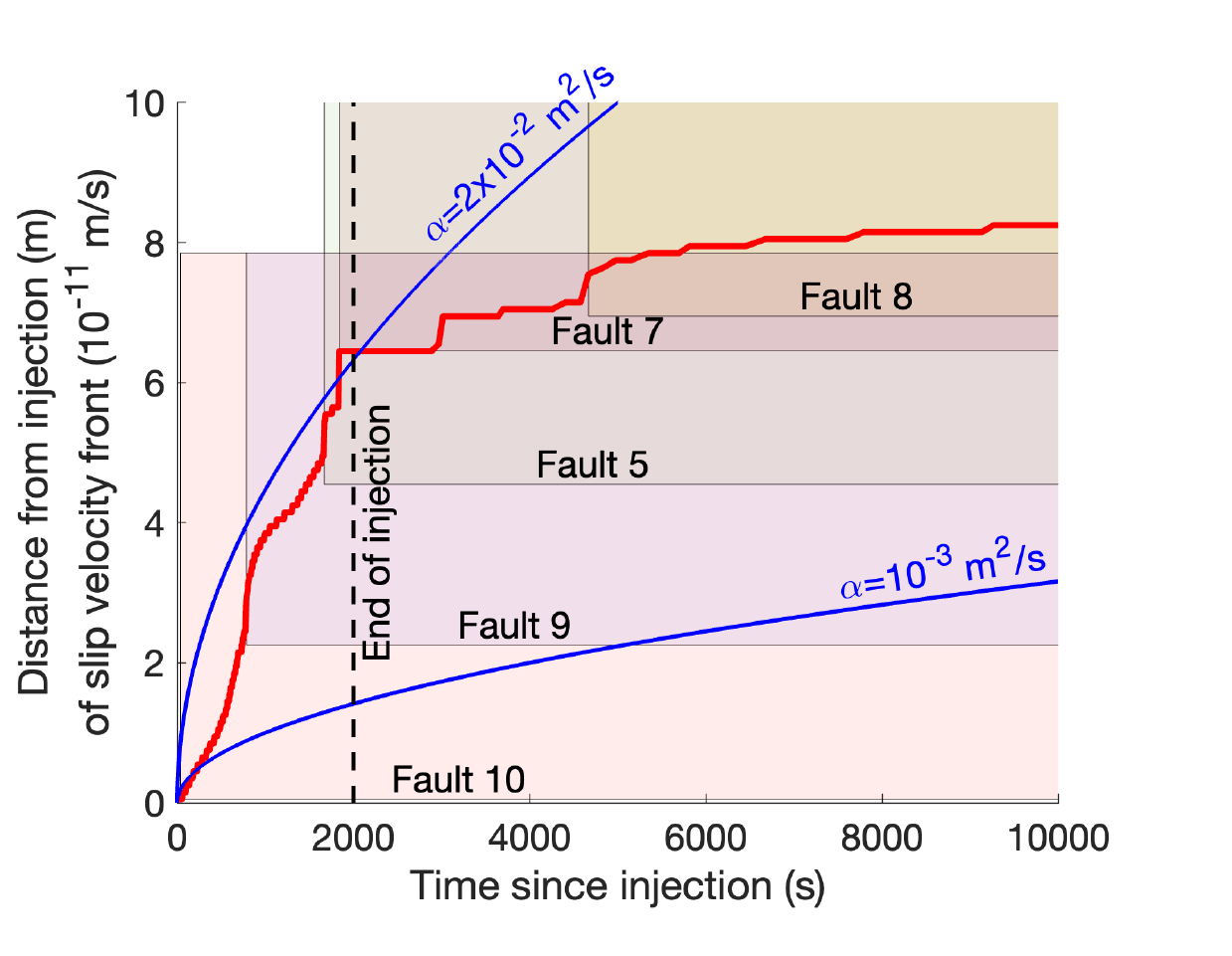}
\caption{Distance from injection point of the slip velocity front defined by the threshold $V=10^{-11}\; \mathrm{m\;s}^{-1}$ (red curve). The transparent rectangles indicate the spatial extent of each fault that has reached the velocity threshold and its duration of slip at a velocity above the threshold. The blue curves show the theoretical diffusion scaling for the hydraulic diffusivity assumed in the model ($\alpha=10^{-3}\; \mathrm{m}^2\;\mathrm{s}^{-1}$) and for a 20 times higher diffusivity ($\alpha=2\times10^{-2} \;\mathrm{m}^2\;\mathrm{s}^{-1}$). The dashed line indicates the end of the injection. }
\label{fig:network_diff}
\end{figure}

To better characterize the slow slip propagation, we defined the position of the slow slip front by a slip-velocity threshold ($V=10^{-11}\; \mathrm{m\;s}^{-1}$) and plotted its distance from the injection point (Fig. \ref{fig:network_diff}). The results reveal that slow slip exhibits a diffusion-like behavior, with a front migration consistent with the scaling of equation~\ref{eq:difusion}, spanning multiple faults within the network. The apparent diffusivity of the slow slip front is approximately 20 times greater than the fluid diffusivity. This behavior is qualitatively consistent with the results of \cite{ciardo2023} on a discrete fault network with a constant friction coefficient. Similarly to earlier results, the fault system operates in a ``critically stressed regime". 

This example highlights the complexity of the interaction between fluid diffusion and stress transfer due to slow slip transients. It is important to note that this is a very simplified case scenario and that more sophisticated 3D geometry, heterogeneous frictional properties and injection history may give rise to the level of complexity that is observed in nature.

% \begin{figure}[H]% discretized fault
% \centering
% \includegraphics[width=\textwidth]{./figures/geometry.pdf}
% \caption{Geometry of the fault system. The injection point is shown by a black dot. }
% \label{fig:geometry_network}
% \end{figure}

\begin{table}[H]
\label{tab:table_network}
\scalebox{1}{
\begin{tabular}{cll}
  \hline
 \multicolumn{3}{c}{Faul frictional properties} \\
  \hline
$a$&0.005 & Direct effect\\
$b$&0.004 & Evolution effect\\
$D_c$&$1\;\mathrm{\mu m}$ & Critical Distance\\
$f_0$&0.6 & Reference friction\\
$V_0$&$1 \; \mathrm{\mu m\;s}^{-1}$ & Reference velocity\\
\hline
\multicolumn{3}{c}{Elastic properties of rock} \\
\hline 
$\mu$&$32.040 \; \mathrm{GPa} $& Shear modulus\\
$\nu$& 0.1541& Poisson's ratio \\
\hline
\multicolumn{3}{c}{Fault hydraulic and fluid properties} \\
\hline
$k$ & $1\times 10^{-15} \;\mathrm{m}^2$ & Permeability\\
$\beta_f$& $9\times 10^{-9}\;\mathrm{Pa}^{-1}$ & Compressibility of the fluid \\
$\beta_{\phi}$&$ 1\times 10^{-9}\;\mathrm{Pa}^{-1}$& Compressibility of the pore\\
$\phi$&$0.1$ & Porosity \\  
$\eta$& $1\times 10^{-3}\;\mathrm{Pa.s}$& Dynamic viscosity \\
$q$& $2\times 10^{-6}\; \mathrm{m\;s}^{-1}$& Flow rate of injection \\
$t_{\text{inj}}$& $10 \text{ years}$& Beginning of injection \\
$\Delta t_{\text{inj}}$& $2000\; \mathrm{s}$ & Duration of injection \\
\hline
\multicolumn{3}{c}{Initial values for mechanical conditions} \\
\hline
$V_{\text{ini}}$&$1 \times 10^{-12}\;\mathrm{m\;s}^{-1}$ & Initial slip velocity\\
$\theta_{\text{ini}}$& $10^6 \; \mathrm{s} $ & Initial state\\
$\sigma_{\text{load}}$  & $5\;\mathrm{MPa}$  & Normal stress\\
\hline
\multicolumn{3}{c}{Length scales} \\
\hline
$ds$&$10 \;\mathrm{cm}$ & Grid size \\  
$L_b$& $1.67 \; \mathrm{m}$& Process zone size
\end{tabular}}
\caption{Parameter values for the simulation of fluid diffusion and aseismic slip propagation along a fault network.}
\end{table}

\newpage

\section{Conclusion}

We presented a robust numerical framework to simulate seismic and aseismic slip on non-planar faults and fault networks, specifically induced by fluid injection and diffusion along faults. Our method seamlessly integrates a boundary element method, tailored to model slip on faults governed by rate-and-state friction, with a finite volume method for accurately capturing fluid diffusion along fault networks. While our 2D model assumes constant fault permeability and omits poroelastic effects in the surrounding rock matrix, it effectively captures the essential mechanisms driving fault slip and seismicity.

To demonstrate the versatility and capabilities of our approach, we analyzed two representative scenarios: (1) fluid injection triggering slow slip on a primary, rough, rate-strengthening fault, which in turn initiates microseismicity on nearby secondary faults, and (2) fluid injection on one fault within a network of intersecting faults, leading to fluid diffusion and slip reactivation across the network. In both cases, the simulated slow slip propagated more rapidly than the fluid pressure diffusion front, revealing complex interactions between fluid-induced slip and the surrounding fault mechanics.
Our results underscore that migration patterns of microseismicity (in the first scenario) and slow slip (in the second scenario) in networks of faults exhibit characteristics resembling a diffusion-like process, but with effective diffusivity values that are very different than the hydraulic diffusivity along faults. These findings reinforce the concept that the observed microseismicity front should not be interpreted as a direct proxy for the fluid diffusion front \citep{wynantsMorel2020,deBarros2021}. Consequently, care must be taken when using microseismic observations to infer fluid diffusivity, as they may reflect mechanical processes distinct from fluid transport dynamics in networks of faults.

This work provides a new simulator for exploring the interplay between fluid injection, fault mechanics, and seismic activity, offering valuable insights for understanding and mitigating induced seismicity in crustal reservoirs. Future developments will aim to incorporate variable fault permeability and poroelastic coupling to extend the model’s applicability to more complex 3D geological settings.

%\section*{Author contribution statement}

\section*{Acknowledgments}

P.R. thanks Barnaby Fryer for insights on the Finite Volume Method. This study was supported by European Research Council (ERC) Starting Grant 101040600 (HYQUAKE).
J.P.A. was supported by the French government through the UCAJEDI Investments in the Future project (ANR-15-IDEX-01) managed by the National Research Agency (ANR).

\appendix
% \section{Table of parameters}
% \comment{Double check table of parameters used in the paper}

% \begin{table}[H]
% \scalebox{1}{
% \begin{tabular}{c|l}
% Symbol & Description \\
%   \hline
%  $\tau_{\text{el}}$& Elastic shear traction due to slip distribution \\
%  $\Delta u$ & Shear slip (mode II) \\
% $\tau_{\text{load}}$& Shear traction loading on the fault\\
% $\mu$ & Shear modulus \\
% $\beta$ & Shear-wave speed \\
%  $V$ & Slip velocity\\
% $\sigma_{\text{el}}$& Elastic normal traction due to slip distribution \\
% $\sigma_{\text{load}}$& Normal traction loading on the fault \\
% $p$ & Fluid pressure \\
% $f_0$ & Friction coefficient at specific velocity $V_0$ \\ 
% $V_0$ & Reference velocity for rate-and-state friction \\
% $a$ & Direct effect in rate-and-state friction \\ 
% $b$ & Evolution effect in rate-and-state friction \\ 
% $\theta$ & State evolution variable \\ 
% $D_c$ & Critical slip distance in rate-and-state friction \\

%  \end{tabular}}
% \caption{Table of parameters}
% \end{table}

\section{Benchmark of the fluid pressure diffusion}
We benchmarked the code using the SCEC SEAS benchmark BP6 of fluid injection at constant flow rate on a planar velocity-strengthening rate-and-state fault with the aging law \citep{lambert2024}. The parameters of the benchmark (see Tab. \ref{tab:table_benchmark}), as well as a more complete description of it can be found on the SCEC SEAS website: \url{https://strike.scec.org/cvws/seas/benchmark\_descriptions.html}. Contrary to the two examples described in the main text, the benchmark problem involves out-of-plane slip (mode III).

\begin{table}[H]
\label{tab:table_benchmark}
\scalebox{1}{
\begin{tabular}{cll}
  \hline
 \multicolumn{3}{c}{Frictional fault parameters} \\
  \hline
$a$&$0.007$ & Direct effect\\
$b$&$0.005$ & Evolution effect\\
$D_c$&$4\;\mathrm{mm}$ & Critical Distance\\
$f_0$&$0.6$ & Reference friction\\
$V_0$&$10^{-6}\;\mathrm{m\;s}^{-1}$ & Reference velocity\\
\hline
\multicolumn{3}{c}{Mechanical properties for rocks} \\
\hline 
$\mu$&$32.04\;\mathrm{GPa} $& Shear modulus\\
$c_s$& $3.464 \;\mathrm{km\;s}^{-1}$ & Shear wave speed \\
\hline
\multicolumn{3}{c}{Fault hydraulic and fluid properties} \\
\hline
$\alpha$ & $0.1\;\mathrm{m}^2\;\mathrm{s}^{-1}$ & Hydraulic diffusivity\\
$k$ & $10^{-13}\;\mathrm{m}^2$ & Permeability\\
$\beta_f+\beta_{\phi}$&$10^{-8}\; \mathrm{Pa}^{-1}$ & Pore and fluid compressibility \\
$\phi$&$0.1$ & Porosity \\  
$\eta$&$10^{-3}\;\mathrm{Pa.s}$ & Dynamic viscosity \\
$q$& $1.25 \times 10^{-6} \;\mathrm{m\;s}^{-1}$& Flow rate of injection \\
$t_{\text{inj}}$&$0\;\mathrm{s}$& Beginning of injection \\
$\Delta t_{\text{inj}}$& $100 \;\mathrm{days}$& Duration of injection \\
\hline
\multicolumn{3}{c}{Initial values for mechanical conditions} \\
\hline
$V_{\text{ini}}$&$10^{-12}\;\mathrm{m\;s}^{-1}$ & Initial slip velocity\\
$\theta_{\text{ini}}$& $ \; \mathrm{s} $ & Initial state\\
$\sigma_{\text{load}}$  & $50\;\mathrm{MPa}$  & Normal stress\\
\hline
\multicolumn{3}{c}{Lengthscales} \\
\hline
$ds$& $10\;\mathrm{m}$ & Grid size \\  
$L_b$&$513\;\mathrm{m}$ & Process zone size
\end{tabular}}
\caption{Table of parameters for the benchmark SCEC SEAS BP6 with the aging law \citep{lambert2024}.}
\end{table}

\begin{figure}[H]% discretized fault
\centering
\includegraphics[width=\textwidth]{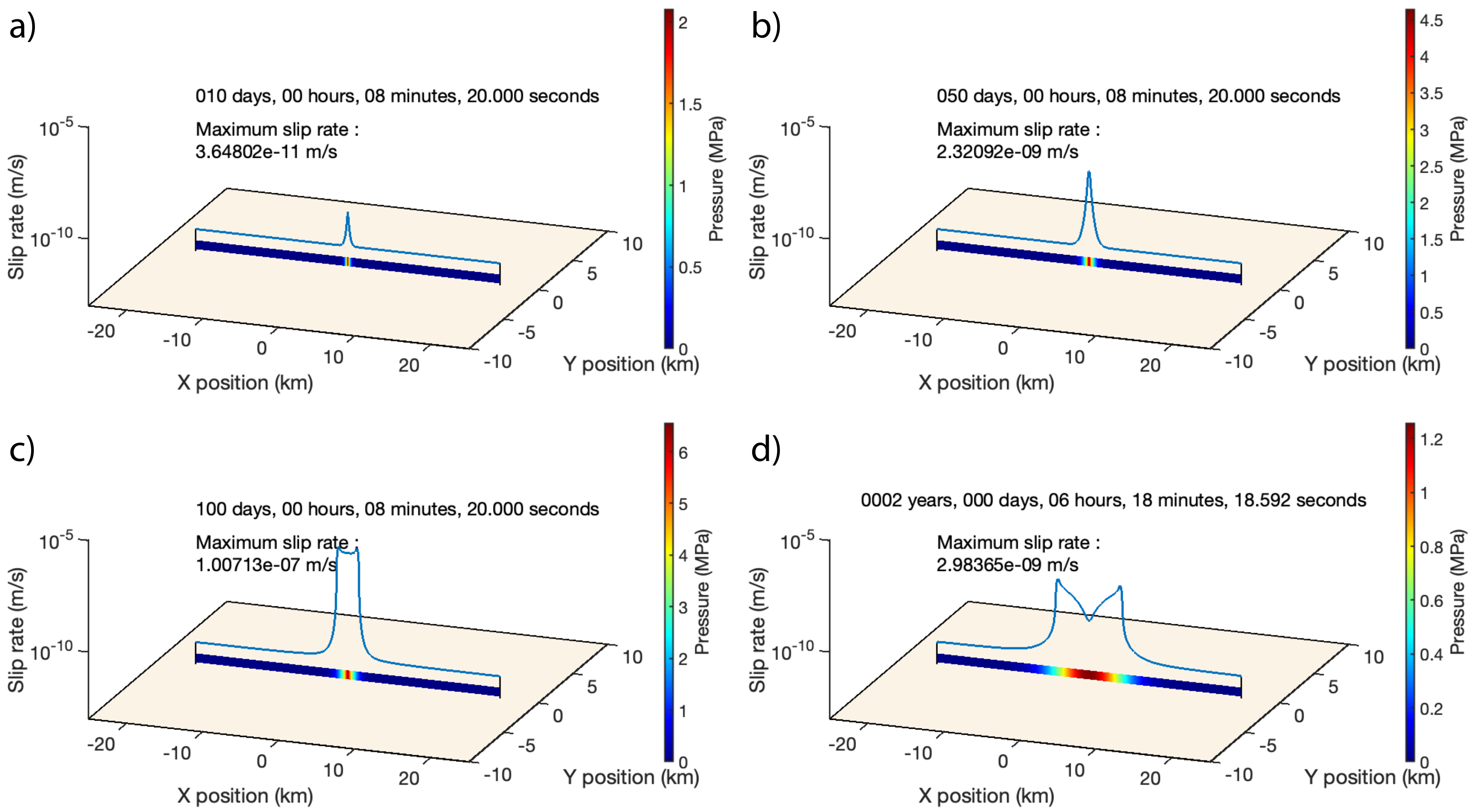}
\caption{Snapshots of the SCEC SEAS BP6 benchmark simulation. Selected times, relative to the beginning of injection, are: (a) 10 days, (b) 50 days, (c) 100 days (this is when injection stops)), and (d) 2 years.}
\label{benchmark1}
\end{figure}
\begin{figure}[H]% discretized fault
\centering
\includegraphics[width=\textwidth]{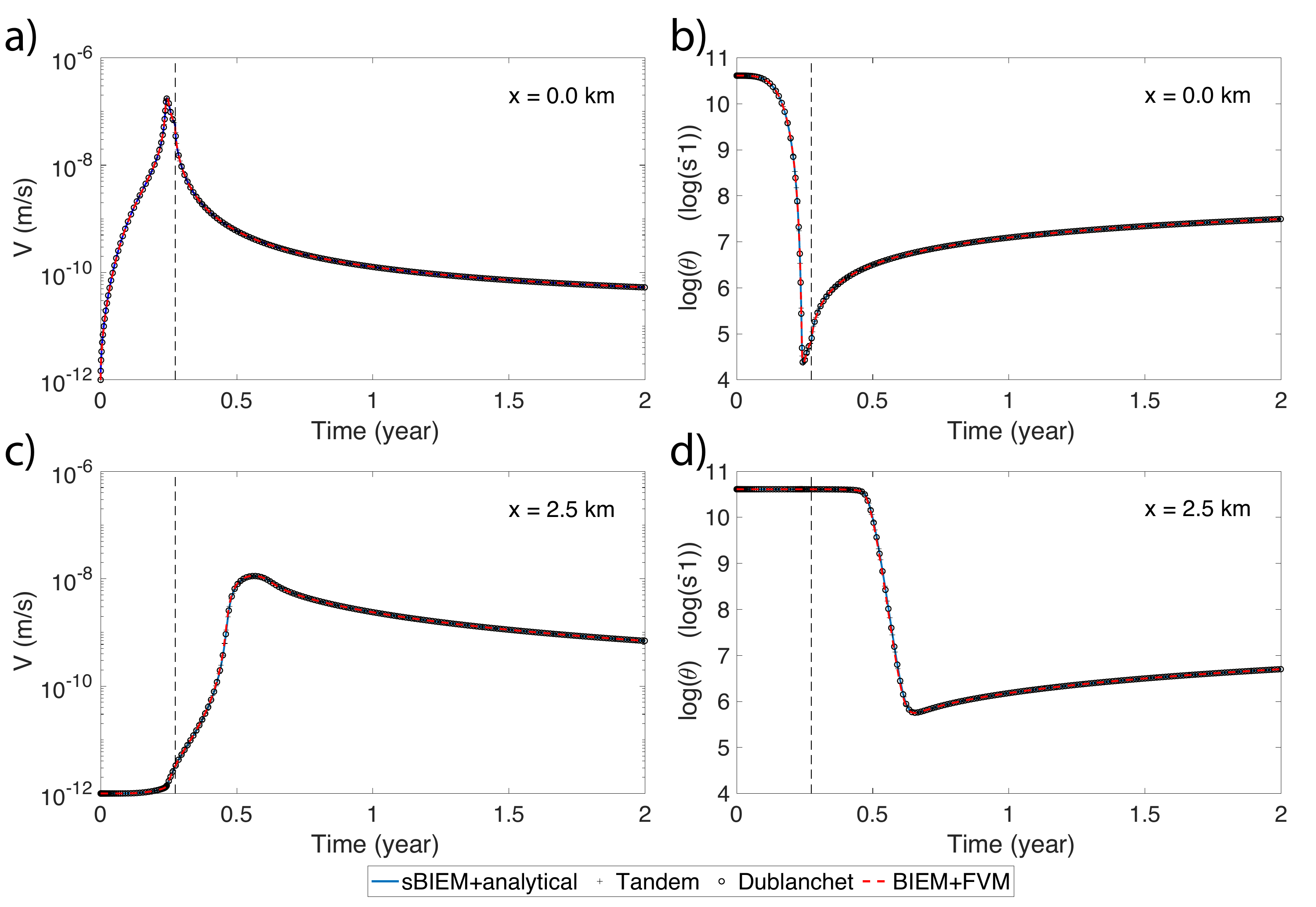}
\caption{(a) Slip velocity at the center of the fault ($x=0.0 \;\mathrm{km}$). (b) State evolution at the center of fault ($x=0.0 \;\mathrm{km}$). (c) Slip velocity at the position $x=2.5 \;\mathrm{km}$. (d) State evolution at the position $x=2.5 \;\mathrm{km}$. The results of our FVM-BEM mthod are compared to those of a finite element method \citep{uphoff2023} and two spectral-time boundary element methods \citep{dublanchet2019,romanet2022}. The dashed line represents the end of the injection.}
\label{fig:benchmark2}
\end{figure}

Fig. \ref{benchmark1} shows different snapshots of the fluid diffusion along the fault, and the triggered slow-slip that follows it. We have compared our new numerical scheme against three other codes: a spectral BIEM code coupled with an analytical solution for the fluid diffusion \citep{romanet2022}, the TANDEM sofware based on the Finite Element Method \citep{uphoff2023}, and another spectral-time BEM method \citep{dublanchet2019}. The results are shown in Fig. \ref{fig:benchmark2}. Our results closely align with the results from other groups both for the slip velocity (Figs. \ref{fig:benchmark2}.a and \ref{fig:benchmark2}.c) and for the state variable (Figs. \ref{fig:benchmark2}.b and \ref{fig:benchmark2}.d).

\begin{figure}[H]% discretized fault
\centering
\includegraphics[width=0.7\textwidth]{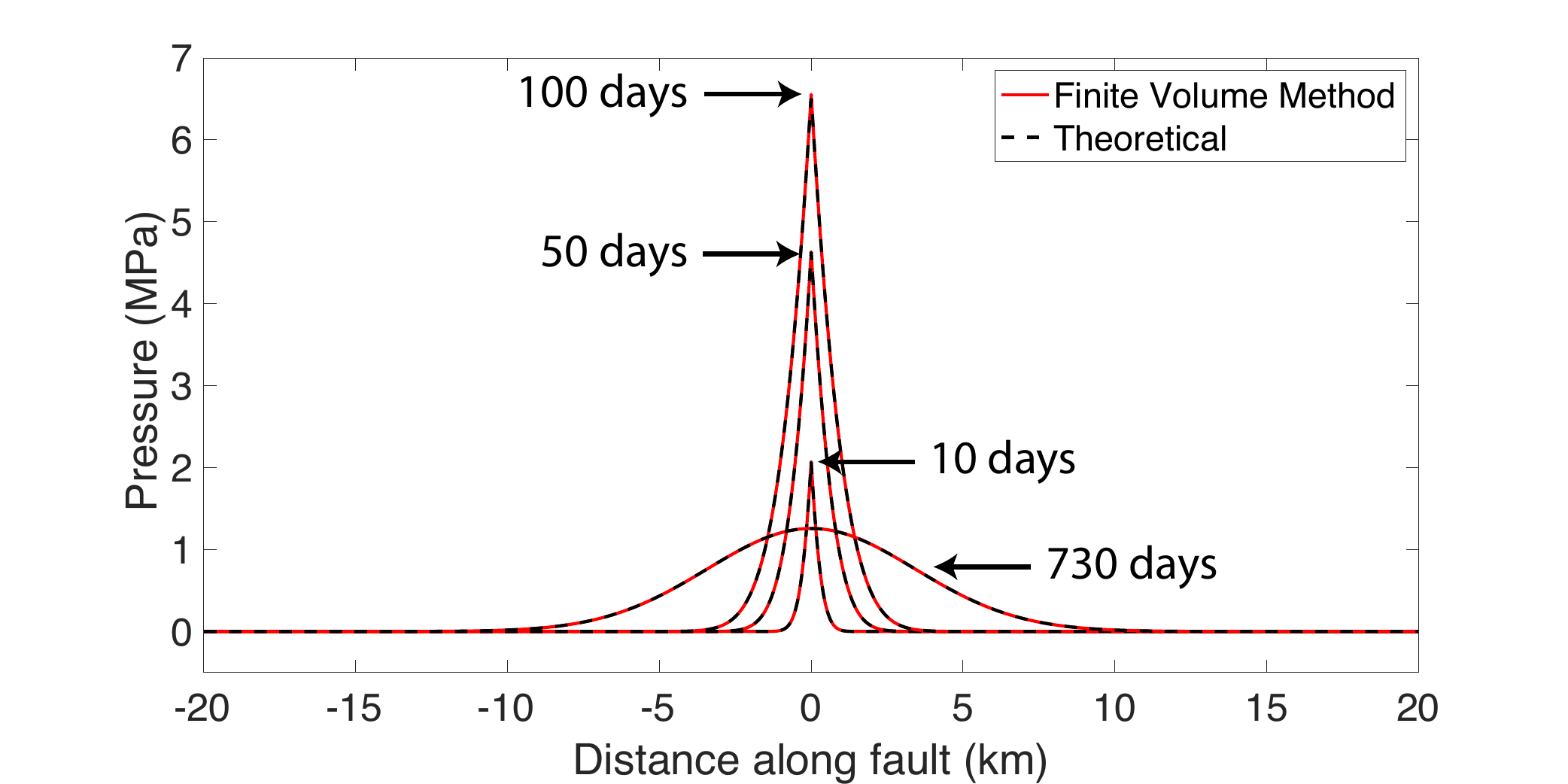}
\caption{Comparison of pressure profiles computed by the Finite Volume Method (FVM) and the analytical solution of the fluid diffusion equation under injection at constant flow rate at $x=0$. }
\label{benchmark3}
\end{figure}

Finally, we also tested the FVM method by comparing the results against the analytical solution for pore pressure diffusion (Fig. \ref{benchmark3}). The results of our FVM closely match with the analytical solution. 

\section{Kernels for BEM}

The kernels relating slip to elastic shear stress $\tau_{\text{el}}(\mathbf{x})$ and normal stress $\sigma_{\text{el}}(\mathbf{x})$ are \citep{tada1997, romanet2020,romanet2024}:
\label{sec:kernels}
\begin{equation}
\begin{split}
K^t_{\text{grad}}(\mathbf{x},\mathbf{y}) &=\left[ 4n_1(\mathbf{x})n_2(\mathbf{x})\gamma_1(\mathbf{x},\mathbf{y})\gamma_2(\mathbf{x},\mathbf{y})+(n_2^2(\mathbf{x})-n_1^2(\mathbf{x}))(\gamma_2^2(\mathbf{x},\mathbf{y})-\gamma_1^2(\mathbf{x},\mathbf{y}))\right] \\
&\times\left(n_2(\mathbf{y})\frac{\gamma_1(\mathbf{x},\mathbf{y})}{r(\mathbf{x},\mathbf{y})}-n_1(\mathbf{y})\frac{\gamma_2(\mathbf{x},\mathbf{y})}{r(\mathbf{x},\mathbf{y})}\right),\\
K^t_{\text{curv}}(\mathbf{x},\mathbf{y}) &=\left[ 4n_1(\mathbf{x})n_2(\mathbf{x})\gamma_1(\mathbf{x},\mathbf{y})\gamma_2(\mathbf{x},\mathbf{y})+(n_2^2(\mathbf{x})-n_1^2(\mathbf{x}))(\gamma_2^2(\mathbf{x},\mathbf{y})-\gamma_1^2(\mathbf{x},\mathbf{y}))\right] \\
&\times\left(-n_1(\mathbf{y})\frac{\gamma_1(\mathbf{x},\mathbf{y})}{r(\mathbf{x},\mathbf{y})}-n_2(\mathbf{y})\frac{\gamma_2(\mathbf{x},\mathbf{y})}{r(\mathbf{x},\mathbf{y})}\right), \\
K^n_{\text{grad}}(\mathbf{x},\mathbf{y}) &=-\left(n_1(\mathbf{y})\frac{\gamma_1(\mathbf{x},\mathbf{y})}{r(\mathbf{x},\mathbf{y})}+n_2(\mathbf{y})\frac{\gamma_2(\mathbf{x},\mathbf{y})}{r(\mathbf{x},\mathbf{y})}\right. \\
&+[2n_1(\mathbf{x})n_2(\mathbf{x})(\gamma_2^2(\mathbf{x},\mathbf{y})-\gamma_1^2(\mathbf{x},\mathbf{y}))-2\gamma_1(\mathbf{x},\mathbf{y})\gamma_2(\mathbf{x},\mathbf{y})(n_2^2(\mathbf{x})-n_1^2(\mathbf{x}))] \\
&\left.\times \left(n_2(\mathbf{y})\frac{\gamma_1(\mathbf{x},\mathbf{y})}{r(\mathbf{x},\mathbf{y})}-n_1(\mathbf{y})\frac{\gamma_2(\mathbf{x},\mathbf{y})}{r(\mathbf{x},\mathbf{y})} \right)\right),\\
K^n_{\text{curv}}(\mathbf{x},\mathbf{y}) &=- \left(n_2(\mathbf{y})\frac{\gamma_1(\mathbf{x},\mathbf{y})}{r(\mathbf{x},\mathbf{y})}-n_1(\mathbf{y})\frac{\gamma_2(\mathbf{x},\mathbf{y})}{r(\mathbf{x},\mathbf{y})}\right. \\
&+[2n_1(\mathbf{x})n_2(\mathbf{x})(\gamma_2^2(\mathbf{x},\mathbf{y})-\gamma_1^2(\mathbf{x},\mathbf{y}))-2\gamma_1(\mathbf{x},\mathbf{y})\gamma_2(\mathbf{x},\mathbf{y})(n_2^2(\mathbf{x})-n_1^2(\mathbf{x}))] \\
&\left.\times \left(-n_1(\mathbf{y})\frac{\gamma_1(\mathbf{x},\mathbf{y})}{r(\mathbf{x},\mathbf{y})}-n_2(\mathbf{y})\frac{\gamma_2(\mathbf{x},\mathbf{y})}{r(\mathbf{x},\mathbf{y})} \right)\right)
\end{split}
\end{equation}
where $\gamma_1(\mathbf{x},\mathbf{y}) = \frac{x_1-y_1}{r(\mathbf{x},\mathbf{y})}$, $\gamma_2(\mathbf{x},\mathbf{y})  = \frac{x_2-y_2}{r(\mathbf{x},\mathbf{y})}$, and $r(\mathbf{x},\mathbf{y}) = \sqrt{(x_1-y_1)^2+(x_2-y_2)^2}$ is the distance between the two points $\mathbf{x}$ and $\mathbf{y}$.
The convention that compressive stresses are positive has been adopted here, contrary to classic boundary integral methods.

\section*{Availibility statement}
The software used in this study can be downloaded from the following link: -to come-.

\clearpage
\bibliographystyle{apalike}
\bibliography{MasterBibliography}
\newpage

\end{document}